# Remote Laboratories:
# New Technology and Standard Based Architecture

Hcene BENMOHAMED, Arnaud LELEVE, Patrick PREVOT
*ICTT Laboratory / INSA Lyon, France*
name.surname@insa-lyon.fr

## Abstract

*E-Laboratories are important components of e-learning environments, especially in scientific and technical disciplines. First widespread E-Labs consisted in proposing simulations of real systems (virtual labs), as building remote labs (remote control of real systems) was difficult by lack of industrial standards and common protocols. Nowadays, robotics and automation technologies make easier the interfacing of systems with computers. In this frame, many researchers (such as those mentioned in [1]) focus on how to set up such a remote control. But, only a few of them deal with the educational point of view of the problem. This paper outlines our current research and reflection about remote laboratory modelling.*

## 1. Introduction

In engineering education, a key-activity to improve the learning process is hands-on experimentation, carried out by simulation tools or laboratory facilities [2]. A remote laboratory is typically a transfer of classical *in-situ* laboratory (one can find in every scientific and technical disciplines) towards distance learning environments.

Motivations for remote laboratories development are:
- sharing heavy and expensive instruments and equipments between institutions,
- anytime and anywhere lab access,
- porting lab activity to distance learning environments,
- resorting to real systems for illustrations, during on-line courses or virtual classrooms,
- putting students in front of real situations and allowing them: to discover system behaviors, to train at using instruments, to verify scientific theories, …

Remote laboratory conception requires technical, pedagogical and computer science competencies. Due to these requirements, it appears to be more complex than other e-learning contexts such as on-line courses, virtual classrooms, e-projects, role-playing, etc. However, this kind of training is essential for scientific and technical disciplines and fits a real need.

More generally speaking, we call E-Labs (Electronic Laboratories), either R-Labs (Remote Laboratories), which offer remote access to real laboratory equipment and instruments, or V-Labs (Virtual Laboratories), which are based on simulations of real systems or phenomena.

At first, we will present E-Lab research context and more particularly R-Lab typical architectures. We will expose R-Lab and V-Lab differences and common points. A few years ago, we initiated a project whose main goal is to build a complete E-Lab open platform to answer to genericity and re-usability requirements of such a system. We will present how we conceive this system, in its global structure within its environment.

## 2. E-laboratories

### 2.1 E-Lab researches

Various R-Labs and V-Labs are proposed in scientific and technical disciplines. For instance, [3] presents a laboratory combining simulation, animation and device access for automation discipline. For his part, [4] developed a virtual chemical engineering laboratory, which was implemented as a supplement to the regular chemical lab course. Later on, [5] conceived a cockpit for web-based experiment in automation discipline. This platform supported collaborative work and was based on activity theory and CSCL (Computer Supported Collaborative Learning).

First V-Labs were typical specific solutions to given experiments but we noticed an evolution of V-Lab community which went on proposing generic architecture such as M.A.R.S. model [6] or SimQuest environment [7]. Generally speaking, we noticed R-Labs followed a similar evolution as V-Labs. In both cases, education point of view was left aside in first realizations. Sometimes, we can observe that content (including scenarios) is melted down in container (architecture). So, it is impossible or difficult to modify or add parts in scenario without re-programming it: the system is closed It also becomes very difficult to reuse e-content for other similar experiments. Another common observation deals with autonomous E-Lab platforms. A part of their functions could be performed by classical LMS (user management, prerequisites managing scheduling, …).

### 2.2. R-Lab usual architecture

A classical and common technical environment for R-Labs is shown in Fig. 1. Users connect to R-Lab through private or public networks. The whole lab is a set of rooms. Each room contains a set of process (devices) connected to local PC (providing a local control using specific software). This connection can be performed by industrial networks, Data AcQuisition (DAQ) board, a General Purpose Interface Bus (GPIB), Programmable Logical Controller (PLC), …

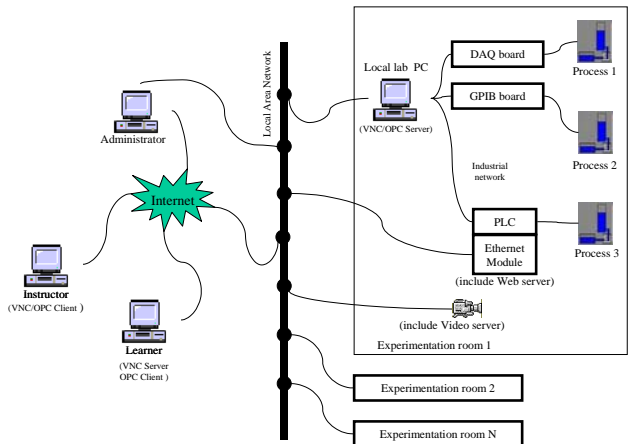

**Fig. 1: Typical E-Lab technical environment**

To obtain a minimum observability, an audio and video feedback is generally necessary. Resorting to augmented reality (a set of extra data super-imposed on a video streaming) as in [8], for instance, favourably compensates for low qualtity webcam streams.

### 2.3. V-Labs vs R-Labs

In our opinion, V-Labs and R-Labs are complementary: specific areas, such as limnology for instance, cannot be directly used for training without dangerously affecting environment. In R-Lab cases, conceivers would create a scale model (a **hard**ware simulator), reproducing same phenomena as in nature, with the advantage to be able to safely and remotely play with. In V-Lab cases, conceivers would propose a theoretical model to be run by a computer (a **soft**ware simulator) with the same previous advantage plus another one: it enables learners to observe hidden or hardly reproducible phenomena. In both cases, learners could train on real or virtual replica and pedagogical scenarios really could be the same ones. Choosing a R-Lab versus a V-Lab implementation is then based on how hard it may be to reproduce at small scale system behavior (homothety does not always work) or to model complex ones.

From an organizational point of view, sharing an experiment is more complex with R-Labs than with V-Labs because with these later, it is simple to create instances of virtual experiments to be used independently and simultaneously [9]. On the contrary, a simultaneous use of a R-Lab experiment requires a time sharing granularity fine enough to enable parallel sessions without having learners to wait too long for an access to the experiment. It also assumes system is able to quickly be set into a given state (i.e. the state which learner left it in last time). This speed of context change is dependant upon experiment nature as magnitude continuity, inertia and thresholds cannot be infringed. In this context, merging R-Labs and V-Labs (RV-Labs) could be a good solution to obtain an optimized time sharing policy; learners could train on simulation while waiting access to real system, for example.

Taking into account advantages and drawbacks of V-Labs vs. R-Labs, it seems to us interesting to develop platforms able to host both of them. We leave the freedom to authors to choose between one sort or the other, according to the context and his pedagogical objectives. In such a platform, learners will have access to both types as in *Colab* project [7].

### 2.4. E-Lab environment

In other respects, E-Labs are ought to be integrated into global educational environments made of Learning Management Systems (LMS) and Learning Content Management Systems (LCMS). In this context, an E-Lab platform should be able to exchange data (user profile, learner evaluation, e-lab scenarios, …), and efficiently use common tools (chat, video, e-content management, …) within such an environment (see Fig. 2).

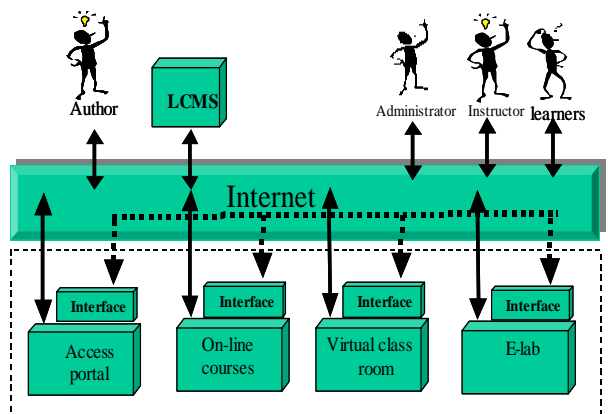

**Fig. 2: E-Lab integration in a global learning environment**

### 2.5. E-Labs at ICTT

A few years ago, we initiated a project whose main goal is to build a complete electronic laboratory platform with an open, extensible and re-usable architecture [10]. To simplify and homogenize notations, let's call it an **ELMS** (Electronic Laboratory Management System). At that time, on the one hand, we noticed a relative lateness in E-Lab development (more especially for R-Labs), compared to other e-learning modes; on the other hand, we needed to develop an E-Lab platform for our emerging open university (called INSA-V). As V-Labs were already studied by many researchers, our first approach consisted in focusing on R-Labs. Still, we keep in mind needs for V-Labs in our conception process in order to conceive an architecture able to host both types

of E-Labs. We set following constraints to ensure its genericity and its integrability; an ELMS:

- should be able to be hosted by an usual LMS, which originally carries out other kind of activity (on-line courses, virtual classrooms, ...),
- should be discipline independent (ought to host physics, chemistry, biology, industrial engineering, automation, ... E-Labs),
- could be extended at any time by plugging-in new devices (water level control, spectrum analyser, robot arm, …) provided by any manufacturer.

Necessary functionalities to run an E-Lab (user management, scenario handling, system control, …) will then be shared between host LMS (typically user profile management, learner evaluation, E-Lab scenario running, …) and ELMS (system setup and control, … In fact, specific functionalities a typical LMS can't provide). So, a typical ELMS consists of four functional environments according to four kinds of actor needs, allowing:

- Authors: to create contents with associated dynamic scenarios, able to be managed by a Learning Content Management System (LCMS)
- Platform administrators: to manage users, schedule sessions, …
- Instructors: to follow, help and evaluate their learners, ...
- Learners: to perform experimentation, to collaborate within teams, to report experimentation production, …

## 3. E-Lab Modelling

Held with several researchers from our laboratory, this work is based on a functional analysis defining which functionalities a remote laboratory platform should propose. Actors are identified by the role they play and by different interactions between them, inside and outside experimentation session. Also, every possible use-case is identified. Fig. 3 gives an idea about the learning process. Learners, with their knowledge, abilities and competencies and, accomplish activities. They use digital, human or hardware resources with the help of instructors, to produce results (an experiment report for example). The pedagogical control here, is an intelligent system to automatically organize learning activities (for both learners and instructor).

Starting from works of [6] and [11] about generic pedagogical simulation tools, we identify five main components communicating with each other by mean of standard protocol (see fig 4, relative to a R-Lab):

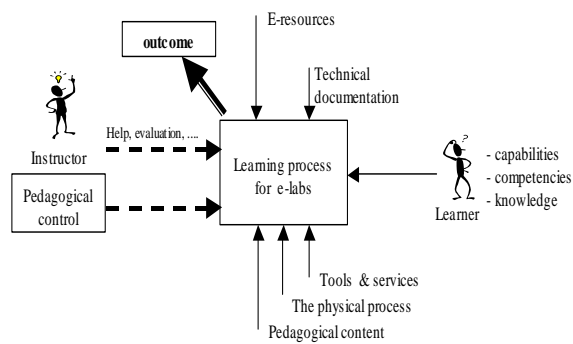

**Fig. 3: Learning process in E-Lab**

- HCI (Human Computer Interface): this central element allows learners to perform experimentation. It communicates with all other components,
- Physical process: features any remotely controlled physical device or process,
- Tools and services: includes all generic and specific necessary tools and services for human communication, collaboration, production, …
- Pedagogical content: consists of content created by authors such as textual or audio instructions, exercises, … in association with scenarios,
- Pedagogical control: is an intelligent control system. It is able to continuously evaluate learners' works and to send in a real time learners' progress to the instructor.

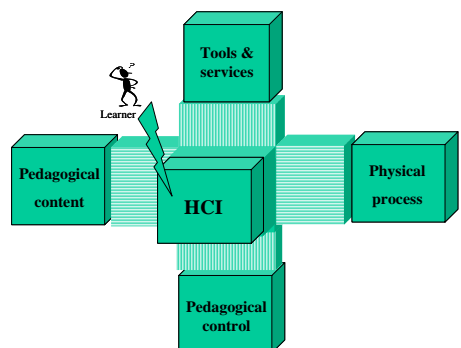

**Fig. 4: Component modelling for a R-Lab**

This approach has the advantage to propose the same global architecture for both remote and virtual laboratories. In V-Lab case, we just replace the *physical process* component by a simulation engine running a system modelling. In §4, we present how we can link this representation to the IMS Learning Design. Follows a brief description of each component.

Tools and services should be shared between host LMS and E-Labs platforms, according to their nature (specific or not to E-Labs). Pedagogical control should be performed by LMS, on the assumption we are able to structure E-Lab scenario as classical e-content. This component will obviously heavily communicate with ELMS (Electronic Laboratory Management system) to send and retrieve remote system or simulation data.

## 4. Technologies and standards for E-Labs

On the one hand, E-Lab platforms are conceived to support experiment devices and instruments, provided by different manufacturers. On the other hand, E-Labs should be integrated in a global learning environment. These needs lead us to study two distinct standards and a research project: the IMS Learning Design for e-learning part (concerning both R-Labs and V-Labs), OPC and OROCOS (both concerning R-Labs) for hardware-software communication (see tab. 1). Next, we give details for each field of use.

| Organisation | IMS | OPC | OROCOS |
|---|---|---|---|
| Developed specifications | - Content Packaging<br>- L I P<br>- Q T I<br>- Metadata<br>- Learning Design<br>- ... | - Data Access<br>- A&E,<br>- XML-DA,<br>- ... | @LAAS<br>@FAW<br>@KTH |
| Studied specification | Learning Design | XML-OPC | @KTH |
| Format | XML | Web services (XML, SOAP, ...) | XML CORBA |
| Interoperability type | E-learning | Hardware | Hardware Software |

**Tab.1: Details of IMS and OPC**

### 4.1. Hardware interoperability

We are splitting our researches about connectivity in two different ways: industrial protocols such as OPC and OROCOS research project.

OPC foundation published several standards and specifications related to the interaction and the communication in process control and manufacturing automation applications. In order to support interoperability between components (physical process, pedagogical control, …), using common protocols is required. OPC protocols are based on Microsoft DCOM (Distributed Component Object Model) technologies. This specification defines a standard set of objects, interfaces and methods. But, as mentioned in [12], DCOM technology presents several drawbacks and limits:
- platform depends on Microsoft environment,
- this protocol was not initially conceived for use over Internet,
- firewalls can create problems for COM message sending.

These problems are solved with the proposition of the XML-DA. This specification creates XML Data schema for use in exposing OPC data to applications over Internet, enabling the sharing of manufacturing information of control devices with applications [12]. Designed for high level and long distance connections, XML-DA is a solution for monitoring over the Internet, accessing data in heterogeneous systems and independent from manufacturers (hardware, software). OPC XML-DA is an interface specification based on XML and SOAP (Simple Object Access protocol) technologies.

Another possible hardware connectivity is studied by the European project OROCOS (Open RObot COntrol Software) [16]. KTH sub-project works on defining a generic device API for an automated system. Adding an abstract layer between automation components and ELMS software should make authoring easier, because authors wouldn't have to worry about low-level programming but would simply deal with high-level services proposed by generic components. This is also interesting for RV-Labs because a virtual system could be controlled the same way as a real one. We are currently digging into this direction to solve generic system control from a scenario and compatibility check.

### 4.2. Integration within an E-learning environment

E-learning specifications, recommendations and standards are proposed by institutions and organizations such as IMS, ADL, AICC, ARIADNE, IEEE, ISO etc. The main goals are:
- to provide a general framework for e-learning architectures,
- to facilitate interoperability between heterogeneous LMS,
- to perform content organization and management,
- to propose Learning Object Metadata and standard frameworks for e-content organization.

In our project, we first studied the use of AICC for E-Labs. We quickly realized that this one presents limitations. It is too much oriented towards e-courses and offers just a simple fixed scenario sequencing, without possibility to freely encode activities for each participants (instructor and learners) and to interface pedagogical part and remote system/simulation part in a standard way. For these reasons we study the possibility of using (or adapting it if necessary) a more modern specification: IMS Learning Design.

Founded in 1997, IMS is a worldwide non-profit organization. Several IMS specifications have become worldwide standards for delivering learning products and services. On February 2003, IMS published the final specification (version 1.0) for Learning Design (IMS-LD). IMS-LD is based on EML [14] (Educational Modelling Language) specification, originally developed by the Open University of Netherlands. IMS adapted it and introduced it into a global set of e-learning oriented specifications. IMS Learning Design specifies:
- a meta- language, designed to support a wide range of pedagogies rather than specific ones, using only one set of learning design and runtime tools,

- learning outcome evaluation,
- single and collaborative learning activities (work team) and interactions between users and role managing,
- tutor helps,
- activities and role managing,
- combined descriptions of the content of a learning resource with descriptions of supported pedagogic activities,
- embedded learning support,
- customized learning scenarios, …

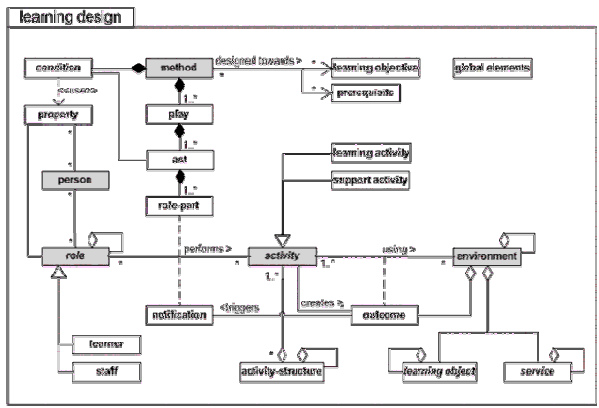

**Fig. 5: IMS-LD (Level C) from [13]**

The three keywords for understanding IMS Learning Design are: *roles*, *activities* and *environment* (Fig. 5). A person put on a role in the teaching-learning process, typically a learner or a staff role. In this role he or she works toward certain outcomes by performing more or less structured learning and/or supporting activities within an environment. The environment consists of appropriate learning objects and services to be used during the performance of the activities. Which role gets which activities at what moment in the process, is determined by the method or by a notification [13].

Comparing our model with the IMS Learning Design, raises following remarks:

- an E-Lab session can be a unit of learning in the IMS Learning Design,
- actors use services and tools to perform E-Lab activities,
- an E-Lab dynamic scenario can be described using method and activity structure,
- E-Lab physical process can be considered as a specific learning object (we can associate with the physical setup an applet such as in [1]),
- our pedagogical control from fig 4. is similar to the IMS Learning Design runtime.

Activities, roles and methods are encoded into an XML file called "*manifest*" (according to the IMS Learning Design Schema). This file along with the additional e-content files are collected into a package (depicted in Fig. 6) to be broadcasted and reused on other similar experiments. So, contents and scenarios can be reused on other compatible systems. One of the advantages of this formalism is the flexibility given to authors to customize their HCI by way of XSLT style sheets. A very interesting aspect in Content Packaging is the ability to slice and recompose packages in order to re-use part or whole scenarios in a standard way (any IMS CP compliant software is ought to do it), such as typical e-content.

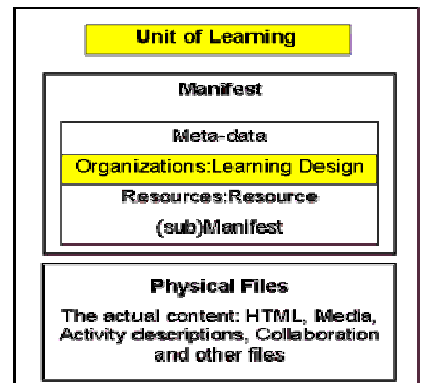

**Fig. 6: Unit of learning from [13]**

At present, the only available runtime environment for IMS Learning Design is CopperCore (since February 2004) [17]. Other projects are in progress such as Learning Design Editor from RELOAD [18] project. CopperCore is a J2EE runtime engine, which can be used to incorporate IMS Learning Design (level A for the moment) in any application(s). One advantage is that it is open source. We are working on using it (with adaptations if necessary) in our operational experimentation platform (presented in details in [1]).

## 5. Conclusion

In comparison with traditional e-contents (virtual classrooms, e-projects, … remote laboratory conception implies additional difficulties. These difficulties are due to system teleoperation requirements, in synchronizing manipulations with e-learning applications, the whole within a standardized platform. This paper gives a global context of electronic, remote and virtual laboratories. Next it presents the current state of our modeling works for an open ELMS which could permit re-using of E-Lab scenarios with the same relative easiness as classical e-content. Concerning E-Lab scenario structure, our works are currently focusing on IMS Learning Design appropriateness.

Moreover, scenario re-using also imposes to check the compatibility between part of or whole scenario and a given (remote or virtual) system. This compatibility check is a future way of study to extend this current work.

These works are will be experimented on our local experimental platform before being tested at a wider scale in our open university (INSA-V).

## 6. References

[1] A. Leleve, H. Benmohamed, and P. Prevot, "Remote Laboratory towards an integrated training system", the *4th proceedings of the International Conference on Information Technology Based Higher Education and Training* (ITHET'03), Marrakech Morocco, July 7-9 2003, pp. 110-115.

[2] S. Ursulet and D. Gillet, "Introducing flexibility in traditional engineering education by providing dedicated on-line experimentation and tutoring resources", *the International Conference on Engineering Education*, Manchester, UK, August 18-21 2002.

[3] B. Wagner, "From Computer-Based Teaching to Virtual Laboratories in Automatic Control", *the 29$^{th}$ ASEE/IEEE Frontiers in Education Conference*, San Juan, Puerto Rico, November 10-13 1999.

[4] C. R. Nippert, "Using Web Based Supplemental Instruction for Chemical Engineering Laboratories", *proceedings of the American Society for Engineering Education Annual Conference and Exposition*, USA, 2001.

[5] Nguyen A.V., Gillet D., Rekik Y., Sire S, "Sustaining the continuity of interaction in web-based experimentation for engineering education", *proceeding of the conference on Computer Aided Learning in Engineering Education (CALIE'04),* February 16-18 2004 pp. 99-110.

[6] J..P. PERNIN, "M.A.R.S. Un modèle opérationnel de conception de simulations pédagogique", PhD thesis, Université Joseph Fourier – Grenoble, January 1996.

[7] T. D. Jong, "Learning complex domains and complex tasks, the promise of simulation based training ", *proceeding of the conference on Computer Aided Learning in Engineering Education (CALIE'04),* February 16-18 2004 pp. 17-23.

[8] D. Gillet, F.Geoffroy, K. Zeramdini, A.V. Nguyen, Y. Rekik, and Y. Piguet, "The Cockpit, An effective Metaphor for Web-based Experimentation in Engineering Education", *the International Journal of Engineering Education (IJEE),* 2003 pp. 389-397.

[9] A. Böhne, N. Faltin and B.Wagner "Self-directed Learning and Tutorial Assistance in Remote Laboratory", *proceedings of the conference on Interactive Computer Aided Learning,* Villach, Australia, September 25-27 2002.

[10] A Leleve, C. Meyer, P. Prevot, "Télé-TP : premiers pas vers une modélisation", *proceedings of the Symposium on Technology of Information and Communication in education for engineering and industry,* Lyon, France November 13-15 2002, pp. 203-211.

[11] G.C. Buitrago, "Simulation et Contrôle Pédagogique : Architectures Logicielles Réutilisables", PhD thesis, Université Joseph Fourier – Grenoble, October 1999.

[12] V. Kapsalis, K. Charatsis, M. Georgoudakis, G. Papadopoulos, "Architecture for Web-based services integration", *The 29th Annual Conference of the IEEE Industrial Electronics Society (IEEE-IECON'03),* Virginia, USA, November 2003, pp. 866-871.

[13] www.imsproject.org
[14] http://eml.ou.nl/eml-ou-nl.htm
[15] www.opcfoundation.org
[16] www.orocos.org
[17] http://coppercore.org/
[18] http://www.reload.ac.uk/